\documentclass[prl,twocolumn,showpacs,preprintnumbers,amsmath,amssymb]{revtex4}

\usepackage{graphicx}
\usepackage{amsmath}
\usepackage{amsbsy}
\usepackage{dcolumn}
\usepackage{bm}

\begin{document}

\title{Andreev states near short-ranged pairing potential impurities}

\author{Brian M. Andersen}
\affiliation{Department of Physics, University of Florida,
Gainesville, Florida 32611-8440, USA}
\author{Ashot Melikyan}
\affiliation{Department of Physics, University of Florida,
Gainesville, Florida 32611-8440, USA}
\author{Tamara S. Nunner}
\affiliation{Department of Physics, University of Florida,
Gainesville, Florida 32611-8440, USA}
\author{P. J. Hirschfeld}
\affiliation{Department of Physics, University of Florida,
Gainesville, Florida 32611-8440, USA}

\date{\today}

\begin{abstract}

We study Andreev states near atomic scale modulations in the
pairing potential in both $s$- and $d$-wave superconductors with
short coherence lengths. For a moderate reduction of the local
gap, the states exist only close to the gap edge. If one allows
for local sign changes of the order parameter, however, resonances
can occur at energies close to the Fermi level. The local density
of states (LDOS) around such pairing potential defects strongly
resembles the patterns observed by tunneling measurements around
Zn impurities in Bi$_2$Sr$_2$CaCu$_2$O$_{8+x}$ (BSCCO). We discuss
how this phase impurity model of the Zn LDOS pattern can be
distinguished from other proposals experimentally.

\end{abstract}


\maketitle

Motivated by the experimental ability to determine the LDOS with
high resolution in both energy and real space using scanning
tunneling microscopy (STM), there has recently been a large
interest in the perturbations caused by impurities in
superconductors. This is because impurities disturb the underlying
superconducting state and hence constitute a natural probe of the
state in which they are embedded\cite{vekhter}. For magnetic
adatoms on the surface of conventional $s$-wave superconductors
such experiments were performed by Yazdani {\sl et
al}\cite{yazdaniscience}. In the superconducting state of BSCCO,
STM measurements have provided detailed information about the
electronic structure near Ni and Zn
impurities\cite{hudson1,hudson2,pannature}. Near Zn it was found
that each impurity generates a sharp conductance peak close to the
Fermi level ($\omega_B \sim -1.5 \mbox{meV}$). By fixing the bias
voltage between the tip and the sample at $\omega_B$, the spatial
structure of this state can be mapped out: it is strongly
localized near the Zn atom with maximum intensity on the impurity
site and second largest intensity on the next-nearest neighbor
sites.

\begin{figure}[b]
\includegraphics[width=8.5cm,height=4cm]{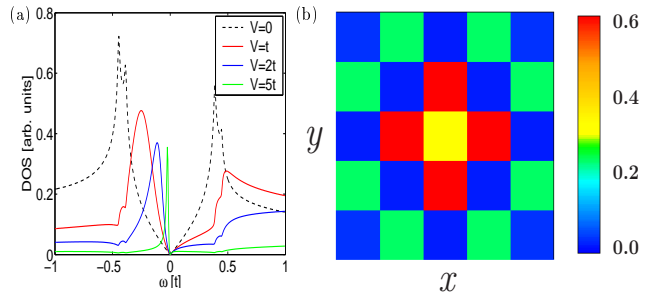}
\caption{(Color online) Left: LDOS in a $d$-wave superconductor at
the impurity site for varying potential scattering strengths $V$
(clean case: dashed curve). As $V$ increases the resonant state
sharpens and moves to lower energy. Right: Real-space LDOS pattern
at the resonant energy $\omega_B$ with a delta-function impurity
potential ($V=5t$) at the center\cite{noteref}.\label{tau3pot}}
\end{figure}

Although there exists a large number of theoretical treatments
dealing with the resonant states generated by nonmagnetic
impurities in $d$-wave superconductors, no consensus has been
reached on their relevance to experiments\cite{vekhter}. Within
the most straightforward scenario where Zn is modelled as a
delta-function potential scatterer, low-energy resonant states are
generated in the unitary limit where the potential is large
compared to all other energy scales of the
problem\cite{rosengren}. For realistic bands with particle-hole
asymmetry, a particular value of the impurity potential $V$ must
be chosen to tune the resonance energy $\omega_B$ to be close to
the Fermi level. Thus, the potential scattering model can
reproduce the energetics of the resonant state. However, as is
well-known, it produces a low-energy spatial LDOS pattern with a
severely suppressed amplitude on the impurity site, in contrast to
experimental results on Zn\cite{pannature}. Physically, this is
clear since a large on-site potential $V$ penalizes any
substantial amplitude of the impurity-state wavefunction on the
impurity site. In fact, within this model the largest intensity is
found on the nearest neighbor site to the impurity. These
properties of the delta-function potential scatterer are shown in
Fig.\ref{tau3pot}.
The discrepancies between experiments and the potential scattering
scenario have led to the suggestion of various filter functions
motivated by the fact that the measurements are performed on the
topmost BiO layer whereas the nonmagnetic impurity substitutes a
Cu atom in the CuO$_2$ plane two layers below the top BiO
layer\cite{zhutunnelfilter,martinfilter,DFT}. The significance of
the filter functions remains controversial. Other scenarios have
focused on the formation of local moments near the Zn
impurities\cite{lee,polkovkondo,ting}. For instance, if the main
effect of nonmagnetic impurities in cuprate superconductors is to
locally break spin singlet bonds, the sharp conductance peak has
been interpreted as a Kondo resonance\cite{polkovkondo}.

While most models assume that an impurity produces a screened
electrostatic potential and possible magnetic exchange effects,
little attention has been paid to the possibility that a defect
can cause local distortions of the interactions which lead to BCS
pairing. We proposed this idea recently in a phenomenological
approach to understanding the correlations between dopant atoms
and gap inhomogeneity in BSCCO\cite{nunner}. In that paper, it was
shown that by assuming that out of plane dopant atoms locally
enhance the pair interaction, many of the peculiar experimental
correlations could be understood\cite{DavisAPS}. Here, we continue
this phenomenological approach, and calculate the Andreev states
associated with atomic scale perturbations in the pairing
potential relevant for superconductors with very short coherence
lengths. This is contrary to the conventional study of Andreev
states within the quasiclassical approach where physical
quantities are assumed to vary slowly on the atomic scale. From
the results for the $d$-wave case, we show that the low-energy
conductance peaks observed near in-plane nonmagnetic impurities in
BSCCO can be explained in terms of localized Andreev states. We
further investigate the robustness of this picture and suggest
some measurements that could be used to test this scenario.

The model is given by the usual BCS Hamiltonian
\begin{equation}
{\mathcal{H}}_0=\sum_{{\mathbf{k}},\sigma} \xi_{\mathbf{k}}
\hat{c}^\dagger_{{\mathbf{k}}\sigma} \hat{c}_{{\mathbf{k}}\sigma}
+ \sum_{{\mathbf{k}}} \left( \Delta_{\mathbf{k}}
\hat{c}^\dagger_{{\mathbf{k}}\uparrow}
\hat{c}^\dagger_{-{\mathbf{k}}\downarrow} + {\rm H.c.} \right),
\end{equation}
where $\xi_{\mathbf{k}}=-2t(\cos k_x + \cos k_y)-4t'\cos k_x\cos
k_y-\mu$ denotes the quasiparticle dispersion with nearest
(next-nearest) neighbor hopping $t$($t'$). For $s$-wave pairing,
$\Delta_{\mathbf{k}}=\Delta_0$, whereas for $d$-wave pairing,
$\Delta_{\mathbf{k}}=\frac{\Delta_0}{2} (\cos k_x - \cos k_y)$. In
terms of the Nambu spinor
$\hat{\psi}^\dagger_{{\mathbf{k}}}=(\hat{c}^\dagger_{{\mathbf{k}}\uparrow},\hat{c}_{-{\mathbf{k}}\downarrow})$,
the corresponding Green's function
${\mathcal{G}}^{0}({\mathbf{k}},i\omega_n)=-\int d\tau
d{\mathbf{r}} \langle T_\tau \hat{\psi}(0,0)
\hat{\psi}^\dagger({\mathbf{r}},\tau) \rangle \exp(i( {\mathbf{k}}
\cdot {\mathbf{r}} - \omega_n \tau))$ in Matsubara representation
is given by
\begin{equation}\label{nambugreensfunctions}
{\mathcal{G}}^{0}({\mathbf{k}},i\omega_n)=\frac{i\omega_n\tau_0+\xi_{\mathbf{k}}\tau_3
+\Delta_{\mathbf{k}}\tau_1}{(i\omega_n)^2-E_{\mathbf{k}}^2},
\end{equation}
where $E_{\mathbf{k}}^2=\xi_{\mathbf{k}}^2+\Delta_{\mathbf{k}}^2$,
and $\tau_i$ denote the Pauli matrices. In real-space, the
perturbation due to delta-function potential impurities in the
diagonal $\tau_3$ channel and modulated pairing in the
off-diagonal $\tau_1$ channel is given by
\begin{equation}
{\mathcal{H}}^{\prime}({\mathbf{r}},{\mathbf{r}}')=\hat{\psi}^\dagger_{{\mathbf{r}}}
\left[ V\delta({\mathbf{r}})\delta({\mathbf{r}}') \tau_3 - \delta
\Delta({\mathbf{r}},{\mathbf{r}}') \tau_1 \right]
\hat{\psi}_{{\mathbf{r}}'}.
\end{equation}
To obtain the resulting LDOS as a function of energy and lattice
sites, one needs to determine the full Green's function
${\mathcal{G}}({\mathbf{r}},i\omega_n)$ given by the Dyson
equation
\begin{equation}\label{dyson}
{\mathcal{G}}({\mathbf{r}},{\mathbf{r}}')={\mathcal{G}}^{0}({\mathbf{r}}-{\mathbf{r}}')+\sum
{\mathcal{G}}({\mathbf{r}},{\mathbf{r}}'')
H^{\prime}({\mathbf{r}}'',{\mathbf{r}}''')
{\mathcal{G}}^{0}({\mathbf{r}}'''-{\mathbf{r}}').
\end{equation}
Thus, by calculating the matrix elements of
\begin{equation}\label{G0realspace}
{\mathcal{G}}^{0}({\mathbf{r}},i\omega_n)= \sum_{\mathbf{k}}
\frac{(i\omega_n\tau_0+\xi_{\mathbf{k}}\tau_3+\Delta_{\mathbf{k}}\tau_1)}{(i\omega_n)^2-E_{\mathbf{k}}^2}
\exp(i {\mathbf{k}} \cdot {\mathbf{r}}),
\end{equation}
the remaining problem is that of a matrix inversion. The solution
is presented in terms of the so-called T-matrix
\begin{equation}\label{tmat}
{\mathcal{G}}({\mathbf{r}},{\mathbf{r}}')={\mathcal{G}}^{0}({\mathbf{r}}-{\mathbf{r}}')+\sum
{\mathcal{G}}^{0}({\mathbf{r}}-{\mathbf{r}}'')
T({\mathbf{r}}'',{\mathbf{r}}''')
{\mathcal{G}}^{0}({\mathbf{r}}'''-{\mathbf{r}}').
\end{equation}
The poles of the T-matrix, or equivalently, the determinant of
$(1-H^{\prime}G^{0})$, determine the bound state energies.

For $s$-wave superconductors, it is well-known that non-magnetic
impurities cannot generate states inside the gap\cite{anderson}.
However, as shown by Yu and Shiba, point-like magnetic impurities
(${\mathcal{H}}^{\prime}({\mathbf{r}},{\mathbf{r}}')=\hat{\psi}^\dagger_{{\mathbf{r}}}
\left[ V_m\delta({\mathbf{r}})\delta({\mathbf{r}}') \tau_0 \right]
\hat{\psi}_{{\mathbf{r}}'}$) of strength $V_m$ produce bound
states at $\omega_B=\pm \Delta_0 (1-\left( V_m \pi N(0)
\right)^2)/(1+\left( V_m \pi N(0) \right)^2)$, where $N(0)$ is the
density of states at the Fermi level in the normal
state\cite{yu,shiba}. Assuming, as a toy-model, a point-like
$\tau_1$ scatterer in an $s$-wave superconductor, a
straightforward calculation shows that an attractive
delta-function $\tau_1$ scatterer generates Andreev bound states
with energies given by $\omega_B=\pm \Delta_0 |(1-(\pi N(0)
\delta\Delta)^2)/(1+(\pi N(0) \delta\Delta)^2)|$. Here,
$\delta\Delta$ is treated as a free parameter and not necessarily
just the self-consistent response of the order parameter to e.g.
an electrostatic potential. A cursory examination of this
expression for the bound state energies $\omega_B$ shows that as
the local gap decreases from the bulk value $\Delta_0$, the
Andreev states are pushed further into the gap. While initially
located near the gap edge, for increased scattering strength the
bound states can in principle be tuned all the way to zero energy,
but this requires $\delta\Delta$ of order the Fermi energy,
outside the framework of conventional weak-coupling BCS theory.


What about $d$-wave superconductors? For a single nonmagnetic
$\tau_3$ delta-function impurity, the T-matrix is diagonal with
$T_{11}=\delta({\mathbf{r}})[V^{-1}-\sum_{\mathbf{k}}
\frac{\omega+\xi_{\mathbf{k}}}{\omega^2-E_{\mathbf{k}}^2}]^{-1}\delta({\mathbf{r}}')$,
$T_{22}=\delta({\mathbf{r}})[-V^{-1}-\sum_{\mathbf{k}}
\frac{\omega-\xi_{\mathbf{k}}}{\omega^2-E_{\mathbf{k}}^2}]^{-1}\delta({\mathbf{r}}')$.
It is the poles of these matrix elements that largely determine
the properties presented in Fig.\ref{tau3pot}. We expect
short-ranged $\tau_1$ scatterers to be more relevant in the case
of $d$-wave cuprate superconductors due to the short coherence
length of these materials. In the $d$-wave case, a point-like
$\tau_1$ impurity is defined as a modulation of the pairing
potential on the four links attached to the impurity site.
Following Shnirman {\sl et al.}\cite{shnirman}, in the case of a
particle-hole symmetric band, one can determine the subgap poles
from the determinant $D(\omega)$ given by
\begin{equation}
D(\omega)=1-\alpha L(\omega) \left( 2 - \alpha L(\omega) + \alpha
\omega P(\omega) \right)\label{det},
\end{equation}
where $\alpha$ parameterizes the local gap suppression, $\delta
\Delta = \alpha \Delta_0$. Here, $P(\omega)$ and $L(\omega)$ are
given by
\begin{equation}
\left( P(\omega),L(\omega) \right) = -\sum_{\mathbf{k}}
\frac{(\omega,\Delta^2_{\mathbf{k}})}{\omega^2-\xi_{\mathbf{k}}^2-\Delta_{\mathbf{k}}^2}.
\end{equation}
We can rewrite Eq. (\ref{det}) in terms of:
$\tilde{P}(\omega)=\tilde{\omega}
K(1/(1-\tilde{\omega}^2)/(\sqrt{1-\tilde{\omega}^2})$,
$\tilde{L}(\omega)=\tilde{\omega}^2/(\sqrt{1-\tilde{\omega}^2})
K(1/(1-\tilde{\omega}^2))+
(\sqrt{1-\tilde{\omega}^2})E(1/(1-\tilde{\omega}^2))$, where
$E(x)$ ($K(x)$) is the complete elliptic integral (of the first
kind), and $\tilde{\omega}=\omega/\Delta_0$ and
$\tilde{\alpha}=4\pi N(0) \Delta_0 \alpha$.
\begin{figure}
\includegraphics[width=8.5cm,height=4cm]{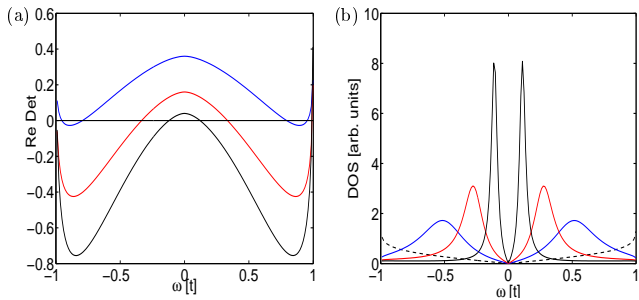}
\caption{(Color online) (a) Real part of the determinant,
$\mbox{Re} D(\omega)$, with $\tilde{\alpha}=0.4$ (top),
$\tilde{\alpha}=0.6$ (middle), and $\tilde{\alpha}=0.8$ (bottom).
(b) LDOS at the impurity site for the same values of
$\tilde{\alpha}$ as in (a), the dashed curve displays the clean
result.\label{detanal}}
\end{figure}
In Fig.\ref{detanal}a we plot the real part of the determinant,
$\mbox{Re} D(\omega)$, for different values of the strength of the
off-diagonal impurity potential. The imaginary part, $\mbox{Im}
D(\omega)$ (not shown), is a monotonically decreasing function as
$\omega \rightarrow 0$ with $\mbox{Im} D(0)=0$. Thus, the
point-like $\tau_1$ impurity in a $d$-wave superconductor also
supports Andreev resonant states. The resulting LDOS at the
impurity site is shown in Fig.\ref{detanal}b: for increased
strength of the gap suppression the Andreev resonance sharpens up
and moves to lower energy. As opposed to the $\tau_3$ point-like
impurity, the resonance exists symmetrically around zero energy
with strongly increased amplitude as $\omega_B \rightarrow 0$. In
general, this symmetry is broken by including e.g. weak $\tau_3$
potentials. One expects the same to be true for more realistic
non-particle-hole symmetric band structures. In this case,
however, it is advantageous to resort to numerics. Since we will
be mostly interested in the case of optimally doped BSCCO, in the
following we fix $t'=-0.3t$, $\Delta_0=0.4t$, and $\mu=-t$
corresponding to $17\%$ hole doping.
\begin{figure}
\includegraphics[width=8.5cm,height=8.0cm]{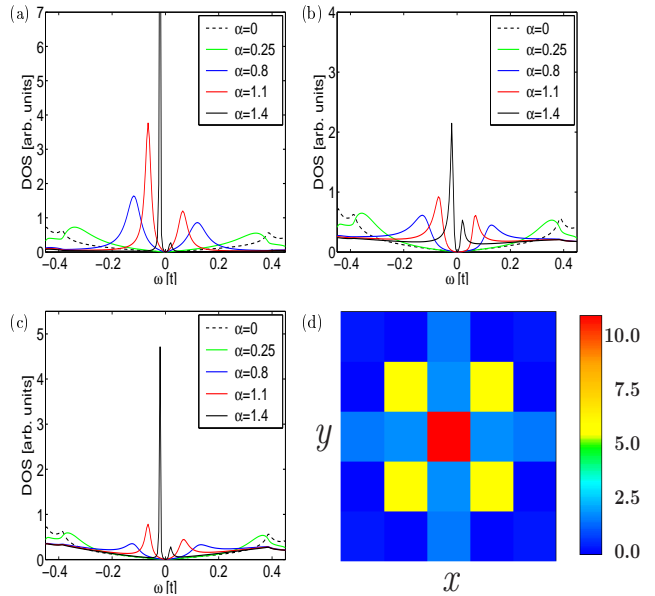}
\caption{(Color online) LDOS for various strengths of the $\tau_1$
scatterer at the impurity site (a), on the nearest neighbor site
(b), and the next-nearest neighbor site (c). In each figure, the
dashed curve shows the bulk LDOS. (d) The real-space LDOS
distribution at the resonance energy $\omega_B$ when
$\alpha=1.4$.\label{tau1tmatrix}}
\end{figure}

In Fig.\ref{tau1tmatrix} we show the LDOS for this band structure
at the three sites nearest to the impurity. Clearly, in this case
the particle-hole symmetry is broken, and the negative bias
resonance dominates the LDOS at the impurity site.
Fig.\ref{tau1tmatrix}(d) shows explicitly that the $\tau_1$
scattering channel produces the {\sl largest maximum on the
impurity site and has a second maximum on the next-nearest
neighbor site}. If we apply the filter function of Ref.
\cite{martinfilter}, the LDOS pattern becomes similar to the one
shown in Fig.\ref{tau3pot}(b) in disagreement with experiments.
When including the $\tau_3$ scattering channel the asymmetry in
Figs.\ref{tau1tmatrix}a,b,c becomes more pronounced. Depending on
the specific band structure, we find that in general the main
features of Fig.\ref{tau1tmatrix} remain valid for $V \lesssim t$.
Therefore, if the conventional argument is eventually proven
correct that a closed unscreened d-shell on Zn necessarily implies
a local potential of several eV, then the present model cannot be
applied to explain the low-energy conductance peak seen near these
impurities in BSCCO\cite{pannature}.

The important point of the results presented here is that
point-like $\tau_1$ scatterers generate well-defined Andreev
states in both $s$- and $d$-wave superconductors. As a specific
example (Fig.\ref{tau1tmatrix}), we found that in order to model
the sharp zero bias conductance peak near Zn impurities in BSCCO,
the local gap has reversed sign compared to its value in the
bulk\cite{note2}. Thus, it is intriguing to consider the idea that
nonmagnetic planar impurities in cuprate superconductors act
mainly as {\sl phase impurities} changing locally the sign of
$\Delta$ on the $x$ and $y$ links compared to the bulk. The
particular value $\delta \Delta_B$ necessary to generate
well-defined low-energy Andreev states near the impurity site
depends on band structure, possible nonzero $\tau_3$ scattering,
and the spatial extent of the pairing modulation. Regarding the
last, we find that for longer ranged phase impurities, the
necessary strength $\delta \Delta_B$ is substantially reduced. For
example, for a $\tau_1$ impurity ranging over two lattice
constants, $\alpha=0.5$ generates LDOS similar to the solid black
curves of Fig.\ref{tau1tmatrix}. Thus, $\delta \Delta_B$ of order
$\Delta_0$ is sufficient to generate well-defined low-energy
states.

Though the approach here is phenomenological and the microscopic
origin of these phase impurities is unknown, they clearly must be
generated by the local perturbations resulting from a Zn ion
replacing a Cu.
In recent work, it was found that the main features of the LDOS
spectra in the inhomogeneous nano-scale regions of BSCCO materials
can be accounted for by local dopant induced modulations of the
pairing interaction\cite{nunner,DavisAPS}. In the present context,
we find that self-consistent solutions of the Bogoliubov-de Gennes
equations generate phase impurities if the pair interaction near
the Zn impurity position is repulsive. Local pairing modulations
can be generated, for instance, by locally modified magnetic
exchanges and/or electron-phonon couplings. A similar approach has
been used previously to explain the enhanced $T_c$ in
twinning-plane superconductors\cite{buzdin}.

\begin{figure}
\includegraphics[width=8.5cm,height=7cm]{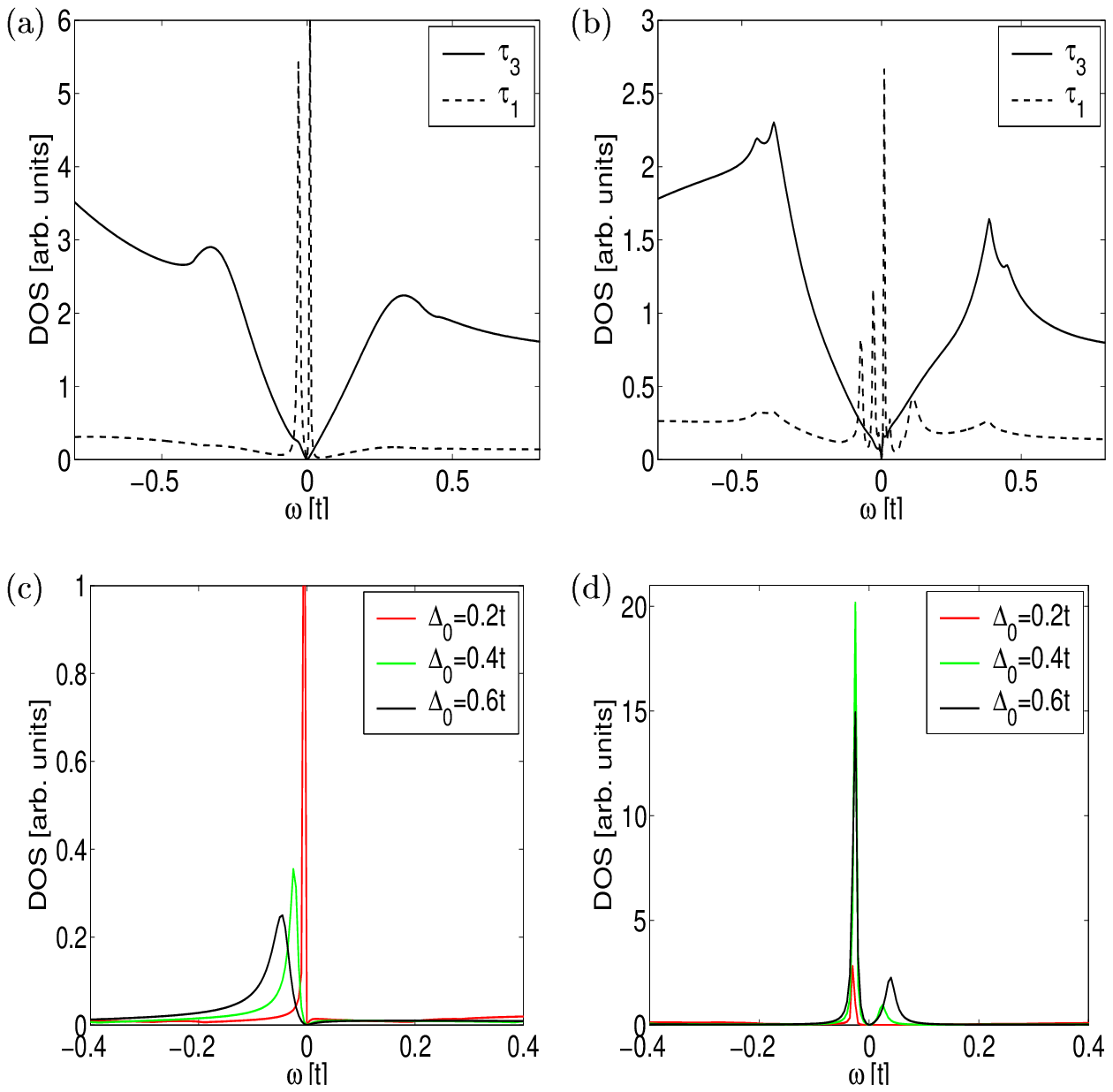}
\caption{(Color online) LDOS at $(0,0)$ (a) and $(2,0)$ (b) when
two impurities are positioned at $(-1,1)$ and $(1,-1)$ for both
the $\tau_3$ (solid) and $\tau_1$ (dashed) channel. In both cases,
the $\tau_3$ LDOS has been rescaled for clarity. (c-d) LDOS at the
impurity site for a single $\tau_3$ (c) and $\tau_1$ (d) scatterer
of fixed strength ($V=5t$ for $\tau_3$, and $\alpha=1.4$ for
$\tau_1$) but varying $\Delta_0$.\label{tests}}
\end{figure}
We end this section by suggesting alternative experimental tests
of whether the $\tau_1$ or $\tau_3$ channel is dominating the LDOS
near Zn in the BSCCO materials. In both scenarios the appearance
of the resonance peak is tied to the superconducting transition
temperature $T_c$. Recently, there has been an increased interest
in the quantum interference caused by nearby impurity
states\cite{morrstravropoulos,andersen1,hirschfeld}.
Figs.\ref{tests}a,b show the resulting LDOS from two interfering
$\tau_3$ or $\tau_1$ impurities. Here, the two impurities are
fixed in close proximity at $(-1,1)$ and $(1,-1)$ and the LDOS is
shown at $(0,0)$ (a) and $(2,0)$ (b). Evidently, there is a
qualitative difference in the resulting interference pattern
between the two scenarios: whereas the $\tau_3$ resonances are
absent, the well-defined Andreev resonances are split and enhanced
by the interference. The latter point remains valid for other
sites (except at the impurity site) in the vicinity of the two
Andreev resonances. Another possible experimental test utilizes
the ubiquitous gap inhomogeneities observed near the surface of
BSCCO\cite{cren,davisinhom1,Kapitulnik1}. Specifically, as shown
in Figs.\ref{tests}c,d the LDOS near Zn impurities in large and
small gap regions is different in the two scenarios: whereas
$\tau_3$ scatterers change both $\omega_B$ and the resonance
amplitude, the $\tau_1$ case exhibits mainly amplitude modulations
with the largest resonance amplitude for intermediate gap
$\Delta_0$.

In summary, we have studied the states near short-ranged
off-diagonal impurities in both $s$- and $d$-wave superconductors.
In both cases we find that the impurities generate low-energy
Andreev states. This offers a new possibility for the origin of
the low-energy conductance peak observed near Zn impurities in
BSCCO .

{\it Acknowledgments.} Supported by ONR grant N(0)0014-04-0060
(BMA,PJH) and Feodor-Lynen Fellowship from the A. v. Humboldt
Foundation (TSN).


\begin{thebibliography}{99}
\bibitem{note2} Note that the bulk $\Delta_0=0.4t$ corresponds to
$\Delta=0.1t$ on each link of the lattice. Thus, for $\alpha=1/4$,
$\Delta=0$ on the links emanating from the impurity site.
\end{thebibliography}
\end{document}